# TOWARDS HIGH-PERFORMANCE NETWORK APPLICATION IDENTIFICATION WITH AGGREGATE-FLOW CACHE

Fei He[1, 2], Fan Xiang[1], Yibo Xue[2,3] and Jun Li[2,3]

[1]Department of Automation, Tsinghua University, Beijing, China
{hefei06,xiangf07}@mails.tsinghua.edu.cn
[2]Research Institute of Information Technology, Tsinghua University, Beijing, China
[3]Tsinghua National Lab for Information Science and Technology, Beijing, China
{yiboxue,junl}@tinghua.edu.cn

## ABSTRACT

*Classifying network traffic according to their application-layer protocols is an important task in modern networks for traffic management and network security. Existing payload-based or statistical methods of application identification cannot meet the demand of both high performance and accurate identification at the same time. We propose an application identification framework that classifies traffic at aggregate-flow level leveraging aggregate-flow cache. A detailed traffic classifier designed based on this framework is illustrated to improve the throughput of payload-based identification methods. We further optimize the classifier by proposing an efficient design of aggregate-flow cache. The cache design employs a frequency-based, recency-aware replacement algorithm based on the analysis of temporal locality of aggregate-flow cache. Experiments on real-world traces show that our traffic classifier with aggregate-flow cache can reduce up to 95% workload of backend identification engine. The proposed cache replacement algorithm outperforms well-known replacement algorithms, and achieves 90% of the optimal performance using only 15% of memory. The throughput of a payload-based identification system, L7-filter [1], is increased by up to 5.1 times by using our traffic classifier design.*

## 1. INTRODUCTION

Modern datacenter and enterprise networks require granular control of traffic to either improve the performance of data transfer or ensure network security. Classifying network traffic according to their application-layer protocols is an essential task for network devices such as next-generation firewall [2], intrusion prevention system, and traffic management system. These devices either perform application specific authentication or quality-of-service assurance based on the application-layer protocol of the traffic.

In early days of the Internet, application identification practices rely to a large extent on the use of transport-layer port numbers. Port-based methods were effective for applications use IANA registered port numbers (for example, HTTP traffic uses port 80 and DNS traffic uses port 25). However, emerging applications such as streaming audio and video, file sharing, and social networks are capable of using non-standard or dynamic ports, encapsulating inside commonly used protocols as a means of evading port-based identification. Therefore, port-based methods can no longer provide sufficient information for network application identification.

Previous works have proposed a number of methods to identify the application associated with a connection. Most of them can be categorized into payload-based methods [3] and statistical methods [4-6]. Payload-based methods are accurate and reliable in most cases, and they are similar with other deep inspection (DI) systems, such as intrusion prevention systems, which inspect the payload of packets searching against a set of signatures. However, these methods require much resources (both computational and memory) since every bytes of packet payloads need to be inspected. Statistical methods usually use machine learning techniques to classify the traffic based on connection properties such as connection duration, packet size distribution per connection, mean packet inter-arrival time and so on [7]. Statistical methods are currently not able to provide fine-grained application identification, and the accuracy of statistical methods is usually not acceptable for in-line network devices. For these reasons, most of production App-ID systems, such as Palo Alto [8], and open source system L7-filter [1], are based on payload-based methods which provide fine-grained and accurate results.

Existing methods cannot meet the requirements of in-line network devices in terms of both throughput and identification accuracy. Payload-based methods are accurate, but they consume a lot of computation power and become bottleneck when hardware resources are limited. Guo et al. [9] presents that the throughput of original L7-filter is around 200Mbps on an latest Intel Xeon platform. Although the throughput of 8-thread parallel version reaches 1.2Gbps, it is far below the wire-speed of modern access networks (e.g. 10Gbps and beyond). Even the throughput of statistical methods is not as good as the common belief. Cascarano et al. [10] demonstrates that a statistical classifier based on Support Vector Machines have comparable computational complexity with a payload-based classifier.

To address the challenges of network application identification, we propose an application identification framework that classifies traffic at aggregate-flow level (defined in Section 3) in this paper. The proposed framework can be used with almost all of the existing methods to improve the overall performance of application identification.

The main contribution of this paper includes:
a) We propose an application identification framework that classifies traffic at aggregate-flow level. Traffic classifier designer can leverage aggregate-flow heuristics provided by the aggregate-flow cache to achieve better identification accuracy or higher throughput.
b) Considering the throughput of traffic classifier as the target, a detailed traffic classifier is designed based on this framework. The traffic classifier caches previous identification results in the aggregate-flow cache to reduce the workload of the backend identification engine. Thus the throughput of the classifier is improved significantly while the identification accuracy is not affected.
c) In order to further optimize the traffic classifier, an effective design of aggregate-flow cache is proposed. We characterize the temporal locality of aggregate-flow cache and propose a frequency based, recency-aware cache replacement algorithm based on the locality analysis.

The rest of the paper is organized as follows. Related work is presented is in Section 2. We describe our aggregate-flow level application identification framework in Section 3. A detailed traffic classifier design for throughput optimization is presented in Section 4. Section 5 demonstrates the design of data structure and replacement algorithm of aggregate-flow cache. Section 6 evaluates the performance of our method. We conclude the paper in Section 6.

## 2. RELATED WORK

Analysis of the application-layer protocol of traffic has always been one of the major interests for network operators. To address the inefficiency of port-based classification, several application identification techniques have been proposed. Generally, they are categorized into payload-based and statistical methods. Payload-based application identification is one of the most used techniques. Sen et al. [3] have proposed a method that use fixed strings and regular expressions to identify P2P applications. Many other studies focus on increase the throughput of payload-based methods. For example, a rich set of research focuses on accelerating regular expression matching [11-13], which is the component consuming most resources in payload-based methods. Guo et al. [9] exploit flow-level parallelism of application identification and utilize multi-core architecture to increase the throughput of application identification system. There are also many identification algorithms based on statistical methods, e.g. machine learning and clustering algorithms. Moore et al. [6] use Bayesian Classifier to perform traffic classification based on per-connection statistics. Li et al. [14] study the method of traffic classification using the SVM. However, all these methods classify traffic on per-connection basis, i.e. only heuristics in each connection are used to perform application identification.

Karagiannis et al. [15] propose an interesting approach that classifies traffic based on the behavior of the hosts generating it. The identification algorithm, called BLINC, analyzes transport-layer information of connections generated by specific hosts, and classifies traffic based on the services the hosts provide or use. This work enlightens us that application identification can be performed beyond connection level. The problem of BLINC is that it cannot classify a single connection, and it cannot perform online classification. The framework proposed in this paper that classifies traffic at aggregate-flow level overcomes these problems.

Several other papers [4, 16, 17] propose running statistical methods and payload-based methods in parallel to improve the accuracy of application identification. In their methods, two or more identification algorithms are run in parallel, and the identification result is combined from the results of each algorithm. Their methods improve the accuracy at the cost of increasing the workload of the identification engine, thus reduce the throughput of the system. This kind of methods can be combined with our aggregate-flow level classification framework to achieve better accuracy without reducing the throughput.

In the next section, we present the proposed application identification framework at aggregate-flow level.

## 3. FRAMEWORK OF APPLICATION IDENTIFICATION AT AGGREGATE-FLOW LEVEL

Most of existing methods only utilize connection-level heuristics. For example, Payload-based methods matches the payload of packets in each connection against a set of pre-defined application signatures. Statistical methods also collects per-connection statistics only, such as packet size distribution per connection, mean packet inter-arrival time, etc. We argue that application identification should be performed above connection-level.

A connection transmitting data between hosts has two endpoints: client endpoint and server endpoint, each of which can be represented by the triple of IP address, transport-layer protocol, and transport-layer port. We define the aggregate-flow as the set of connections with a common server-endpoint. Connections belong to one aggregate-flow are usually generated by the same application at most of times for the following reasons:

- A server endpoint is generally multiplexed between a variety of different connections, which means only one connection needs to be classified and all the other can share the result.
- For most applications, the server endpoint is listened to by the server-side of an application, and the application binding to a specific server endpoint will not change frequently.
- The popularity of various server endpoints could be highly uneven due to the existing of well-known services, e.g. popular websites, mail servers, etc.

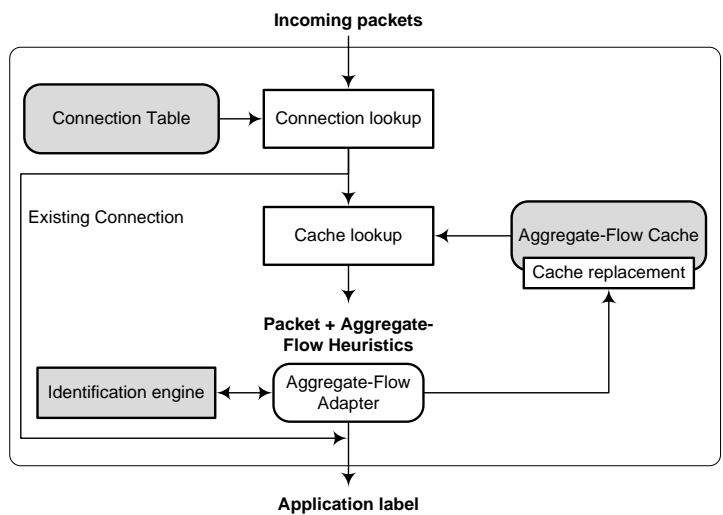

Figure 1. Framework of application identification at aggregate-flow level.

Since connections in each aggregate-flow are generated by the same application, a traffic classifier framework leveraging aggregate-flow level heuristics is proposed in order to achieve better performance than existing methods that identifies at connection-level. As illustrated in Figure 1, the proposed framework consists of a connection table as ordinary traffic classifier. After the application-layer protocol of a connection is identified by the identification engine, it is added to the connection table. The following packets of that connection are not processed by the identification engine. The proposed classifier framework uses a data structure called aggregate-flow cache to capture and store aggregate-flow level information. In more detail, each entry in the aggregate-flow cache consists of a triple (server IP, server port, transport-layer protocol) as the key, an application label field stores previous identification result, and any other aggregate-flow level heuristics needed by the classifier. If a packet finds the corresponding aggregate-flow entry in the cache, the packet along with the aggregate-flow heuristics is sent to the aggregate-flow adapter. The aggregate-flow adapter can be designed to achieve different optimization targets. It determines which packet with what kind of heuristic is sent to the identification engine, and also what kind of identification result should be captured and stored in the cache. For example, if a statistical method is used as the identification engine, the adapter can be designed to collect and store aggregate-flow level statistics in the cache. Both aggregate-flow level statistics and connection level statistics are sent into the identification engine to achieve better identification accuracy. The target of this paper is to improve the throughput of identification engine, a traffic classifier under this framework is presented in the next section to achieve this goal.

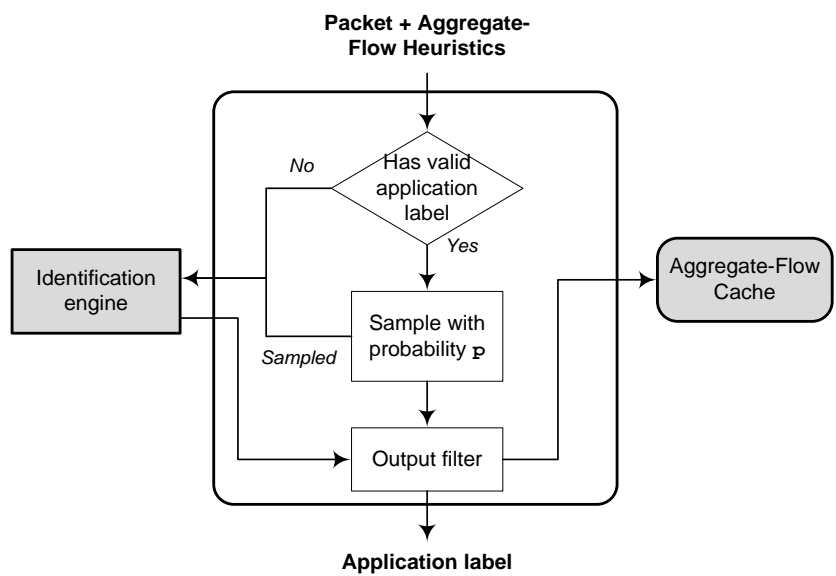

Figure 2. Design of Aggregate-Flow Adapter.

## 4. Design of Traffic Classifier Based on Aggregate-Flow Cache

Since the accuracy and reliability of payload-based methods are acceptable for in-line network devices, it is more urgent to increase the throughput of application identification system. In this section, we demonstrate a traffic classifier designed under the proposed framework to achieve the goal of throughput improvement.

This classifier use a payload-based method as the identification engine. The goal of this design is to reach much better throughput than existing payload-based methods by leveraging the aggregate-flow cache. Figure 2 shows the structure of the aggregate-flow adapter used in this traffic classifier. In this case, the aggregate-flow heuristic is application label of the aggregate-flow the packet belongs to. If a connection of the aggregate-flow has been identified before, the application lable is a valid value. The packets can be marked as the same application label. Only when the packet does not find a match in aggregate-flow cache or the application lable is invalid, the packet is sent to the identification engine for application identification based on its payload. Actually, this traffic classifier perform classification on a per-aggregate-flow basis instead of per-connection basis, which can reduce the workload of the identification engine dramatically.

Although the application of an aggregate-flow generally does not change in the short term, the corresponding cache entry needs to be updated when it changes. To deal with the validation of aggregate-flow entries, the adapter samples each connection that matches a cache entry with a probability. If a connection is sampled, its packets are still sent to the identification engine. The identification result is sent to the output filter module to update existing entries.

The output filter in the adapter also determines whether the identification result of a connection should be cached. There are some special cases that should not be cached: (a) The data connection of applications using dynamic port-negotiation mechanism, e.g. FTP, and SIP, should not be cached, since the server endpoint of this kind of connection is randomly generated and may be used only once. (b) Incoming connections to a proxy server is another category that should not be cached. A proxy server listens at a port that accepting multiple applications. Therefore, connections belongs to the corresponding aggregate-flow may not be generated by the same application.

Storing previous identification result in the aggregate-flow cache can significantly reduce the load of the identification engine. However, when deploying this classifier on network devices, the memory usage of the aggregate-flow cache is an important thing to consider. We propose an efficient aggregate-flow cache design in the next section.

## 5. Efficient Aggregate-Flow Cache Design

To design an efficient aggregate-flow cache, we need to a data structure optimized for both cache lookup and cache replacement operations. In addition, an efficient cache replacement algorithm is very important to get high cache hit ratio using limited memory space. We demonstrate data structure and cache replacement algorithm design in the following subsections.

The maintenance of aggregate-flow cache is different from that of connection table. Concurrent number of connections of a given link is usually within a certain range, since a majority of connections time out quickly. However, it takes much longer time for an aggregate-flow to time out, if defining the timeout of an aggregate-flow as the status that all the connections belong to this aggregate-flow time out. Actually, an aggregate-flow times out only when the server endpoint is closed or changed. The time when a server endpoint is not used is usually not predictable on a network device. Therefore, we use a fixed size memory for the aggregate-flow cache, and do not maintain timeout status for every each aggregate-flow entry. Instead, an aggregate-flow entry is removed only when the cache is full. An efficient cache replacement algorithm is needed for this design to store aggregate-flow entries with high probability of being referenced in the future in the cache. We characterize the temporal locality of aggregate-flow using traffic traces from different networks (detailed information of these traces are shown in Section 6), and derive an algorithm to capture the temporal locality.

Table 1. Value of α for each trace.

| Trace | THU1 | THU2 | LBL1 | LBL2 | CAIDA1 | CAIDA2 |
|---|---|---|---|---|---|---|
| α | 0.90 | 0.98 | 1.26 | 1.19 | 0.69 | 0.73 |

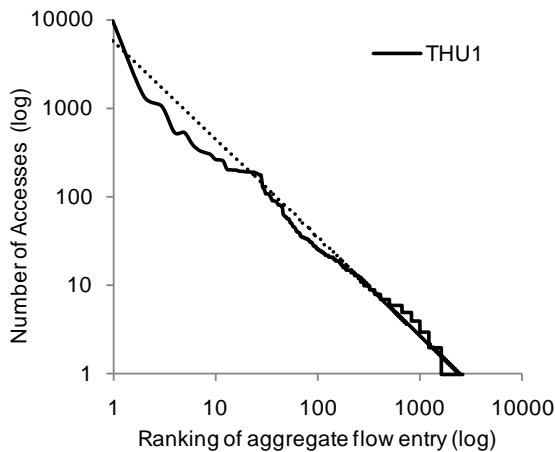

Figure 3. Frequency of aggregate flow accesses versus ranking.

### 5.1 Analysis of temporal locality

Temporal locality means that an entry just referenced has a high probability of being referenced again in the near future. Increased temporal locality generally improves cache performance. Temporal locality of reference in aggregate-flow cache emerges from two distinct phenomena: the long-term popularity of aggregate-flows and the short-term temporal correlations of references. We study the long-term popularity and the short-term temporal correlations separately in order to guide the design of replacement algorithm.

Intuitively, aggregate-flows associated with endpoints of well-known services may contain a larger portion of connections than other aggregate-flows. The log-log scale plot in Figure 3 shows that the popularity of various aggregate-flows is highly uneven. Actually, Zipf's law [4] is applicable to characterize the popularity of aggregate-flows. Zipf's Law states that the popularity of the $n^{th}$ most popular object is proportional to $1/n$. We use the value α in "Zipf-like" distributions as a metric for popularity skewness. In such a distribution:

$$P(O_n) \propto n^{-\alpha}$$

in which $P(O_n)$ is the probability of a reference to the nth most popular object. Table 1 presents a least-square fit of the alpha value of different traffic traces. The experiment shows that long term popularity follows Zipf-like distribution and exhibits different α. Larger α values are observed in edge networks, and the α values of backbone-network traces are smaller.

After identifying an appropriate metric for capturing popularity, it remains to study the correlation component of temporal locality. The stack distance model [Evaluating techniques and storage hierarchies] is a widely used approach to capture and characterize temporal locality. For any given reference to entry $i$, the corresponding stack distance is the number of unique references since the last reference to $i$. However, stack distance cannot distinguish the causes of temporal locality directly.

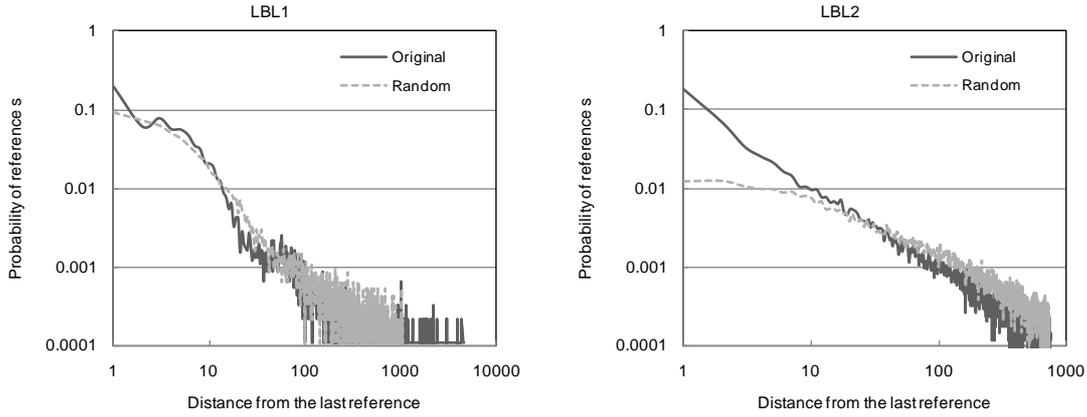

Figure 4. Probability distribution of inter-reference distance of the two traces.

Table 2. Slopes of the curves of the probability distribution of inter-reference distance.

| Trace | THU1 | THU2 | LBL1 | LBL2 | CAIDA1 | CAIDA2 |
|---|---|---|---|---|---|---|
| Original | 0.97 | 0.88 | 0.65 | 1.11 | 0.50 | 0.54 |
| Scrambled | 1.01 | 0.91 | 0.91 | 0.88 | 0.41 | 0.59 |

In order to measure short term popularity separately, we measure the stack distance when the packet streams are subjected to a random permutation. The temporal correlations can be eliminated by applying a random permutation to the packet streams while preserving the popularity distribution. Figure 4 presents the probability distribution of inter-reference distance of two different traces. For trace LBL1, there is no significant change in the distribution between the original trace and the randomized trace, which means the temporal locality in this trace is predominantly determined by the popularity distribution. However, the result of trace LBL2 presented in the right plot of Figure 4, reveals some difference between the original trace and the randomized trace. Table 2 shows the slopes of the curves for different traces using a least-square fit. We can see that temporal correlation only exists in some traces and the intensity of this effect is not clear.

Our characterization of temporal locality indicated that long-term popularity is the predominant and reliable contributor to temporal locality, but that temporal correlation exists in some traces. It advocates us to design an aggregate-flow cache that mainly considers the skewed popularity, but also consider the effect of temporal correlation.

**5.2 Data structure and replacement algorithm**

The basic cache replacement policy that leverages the skewed popularity is Least-Frequently-Used (LFU). LFU infers object popularity directly from the reference history, and replace the entry with the least access frequency when a new entry is added to the full cache. It should be noted that LFU can be implemented in two different forms: perfect LFU and in-cache LFU. Perfect LFU keeps the frequency information of all the objects, including both cached and evicted objects. On the other hand, in-cache LFU only maintains the frequency of cached objects. Perfect LFU is obviously unrealistic due to the large scale of server endpoints, while the performance of in-cache LFU is not good enough since it does not represent all requests in the past.

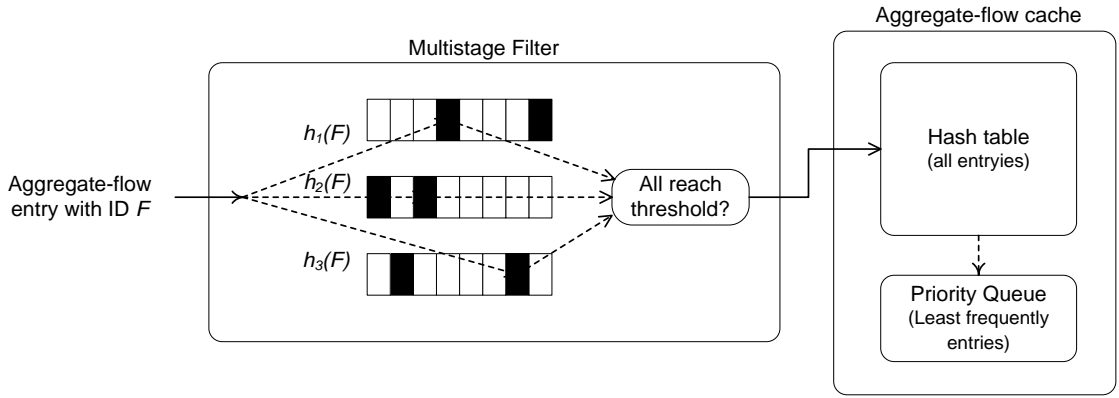

Figure 5. Design of aggregate-flow cache.

In order to capture the most frequently accessed aggregate-flow entries in the cache, and use little extra memory, we employ a data structure called multistage filters [18] to identify frequently accessed aggregate-flows. An aggregate-flow is added to the cache, only if its frequency exceeds a pre-defined threshold *T*.

The basic structure of a multistage filter is shown in the left-side of Figure 5. The building blocks are several hash stages operate in parallel. Each hash stage is a table of counters indexed by a hash function computed on an aggregate-flow ID. All the counters are initialized to 0. Before an aggregate-flow entry is added to the cache, a hash value on its aggregate-flow ID is computed and the corresponding counter is added by one. If the aggregate-flow appears more than threshold *T*, its counter will exceed the threshold. However, since the memory for counters is limited, many aggregate-flows will map to the same counter. This will cause false positives, since several aggregate-flows hashed to the same counter can add up to a number larger than the threshold. To reduce the number of false positives, multiple stages with independent hash functions are used. An aggregate-flow entry is passed to the aggregate-flow cache, only if hashed counters in all stages exceed the threshold. Following the analysis in [18], we use *d* to denote the number of stages, *k* for the stage strength which is the ratio of the threshold and the average size of a counter. The probability of an aggregate-flow of frequency $f \le T(1-1/k)$ passing a *d*-stage filter is at most $P_f \le (\frac{1}{k}\frac{T}{T-f})^d$. The multiple stages reduce the probability of false positives exponentially in the number of stages, which make it possible for us to capture the most frequently accessed aggregate-flows using little extra memory.

Aggregate-flows pass the multistage filter is added to the cache. If the cache is full, the replacement always happens among entries with the lowest frequency. Usually, an implementation of such a policy would store all the entries in a priority queue with access frequency as the key. This results in an expensive *O(nlogn)* time complexity for cache lookup. Fortunately, there are a large number of entries that equal to the lowest frequency, and almost all the replacements happen among these entries. Our implementation uses a hash table to store all the cache entries, and a priority queue to store the entries with the lowest frequency. The priority queue uses the last access time of aggregate-flow as the key. Our implementation selects the oldest one from entries that are least frequently accessed as a candidate for eviction. The implementation needs only *O(1)* time for each cache lookup, and *O(mlogm)* (m is the average number of entries with the lowest frequency) time for each replacement. Experiments in the next section will show that this algorithm outperforms LRU and LFU in almost all the circumstances.

Table 3. General information of our traces.

| Set | Date | Link bandwidth | IP address | Packets | Connections | Aggregate-flows |
|---|---|---|---|---|---|---|
| THU1 | 2005-10-15 | 1Gbps | 19K | 2628K | 36K | 3897 |
| THU2 | 2005-10-15 | 1Gbps | 8.6K | 2455K | 18K | 3484 |
| LBL1 | 2004-12-15 | N/A | 1.0K | 8439K | 9K | 412 |
| LBL2 | 2004-12-15 | N/A | 3.0K | 9257K | 32K | 1597 |
| CAIDA1 | 2009-09-17 | 10Gbps | 1194K | 38485K | 2571K | 487K |
| CAIDA2 | 2010-03-25 | 10Gbps | 1334K | 41083K | 2576K | 525K |

## 6. PERFORMANCE EVALUATION

In this section, we demonstrate the effectiveness of our approach when applied to real world traffic traces. After describe the datasets used and the experiment environment, a proof of concept is given. Then we present the performance of our cache replacement algorithm comparing to several well-known algorithms. Finally, the throughput and memory usage of the proposed traffic classifier is evaluated.

### 6.1 Experiment Setup

We use three sets of traffic traces from academic, enterprise and backbone networks. The first set (THU1, THU2) are traffic traces with packet payload collected on the Internet link of a campus network with about 1000 computers at Tsinghua University. The second set (LBL1, LBL2) are packet header traces of enterprise network, provided by LBNL/ICSI Enterprise Tracing Project [19]. The third set (CAIDA1, CAIDA2) contains packet header traces from CAIDA's monitors on OC192 Internet backbone links [20]. Table 3 lists general information of our data sets.

We implement a trace-driven traffic classifier based on Libnids [21] to test the performance of our design when processing real-world traffic. The backend application identification engine used in the classifier is a widely-deployed payload-based engine, L7-filter [1]. All the experiments were obtained on a Server with a Xeon E5504 CPU (4 cores at 2.0 GHz) and 4 GB DDR3 memory.

### 6.2 Evaluation Results

We first give a proof concept of our method by comparing the number of aggregate-flows with the number of connections. Figure 6 shows that the numbers of aggregate-flows are among 4.7% to 20.4% of the numbers of connections for different network environments. Therefore, the number of connections needs to perform payload-based application identification can be reduced by up to 95%, which increases the performance of the traffic classifier significantly. Actually, the ratio of number of aggregate-flows and number of connections is mainly determined by the applications in network. Since most applications in general networks uses persistent server endpoints, the proposed framework of classifying traffic at aggregate-flow level should be effective.

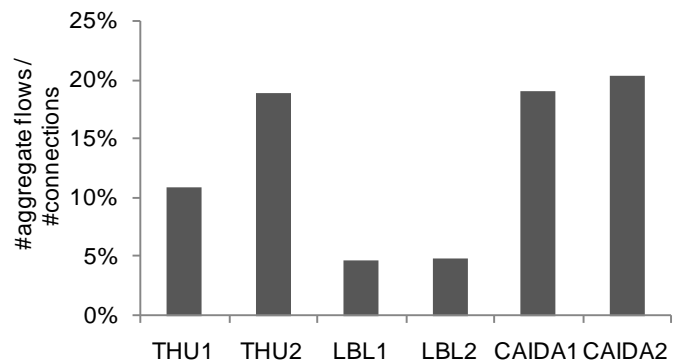

Figure 6. Number of aggregate-flows divided by number of connections.

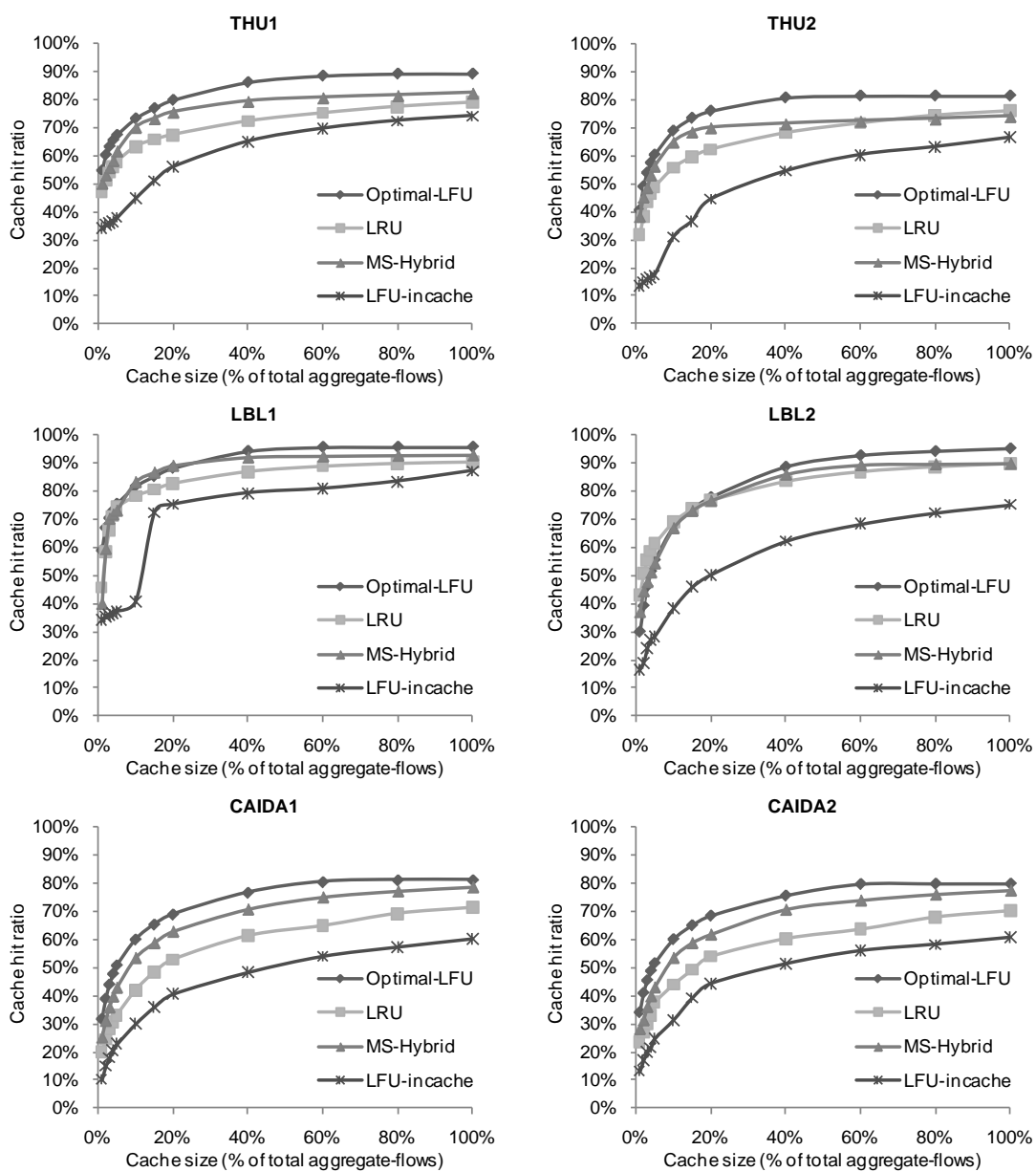

Figure 7. Cache hit ratio of different replacement algoritms

Figure 7 presents the result of our cache replacement algorithm (Multistage-Hybrid, MS-Hybrid) comparing with two well-known replacement algorithm Least Recent Used (LRU) and in-cache LFU. The cache hit ratio of Optimal-LFU is calculated assuming the popularity of every aggregate-flow is known in advance, which is unrealistic. It is the optimal result frequency-based replacement algorithm can reach, and is used as a benchmark for comparison. Our replacement algorithms outperform LRU and in-cache LFU on almost all the traces, especially when the cache size is small. In additional, the cache hit ratio of MS-Hybrid is very close to that of the Optimal-LFU. The result of LRU on trace LBL2 is a little bit better than MS-Hybrid as the analysis in section 5.1 shows that the temporal correlation in LBL2 is intensive. Still, the time complexity of MS-Hybrid is only $O(1)$, smaller than $O(nlogn)$ of LRU and LFU.

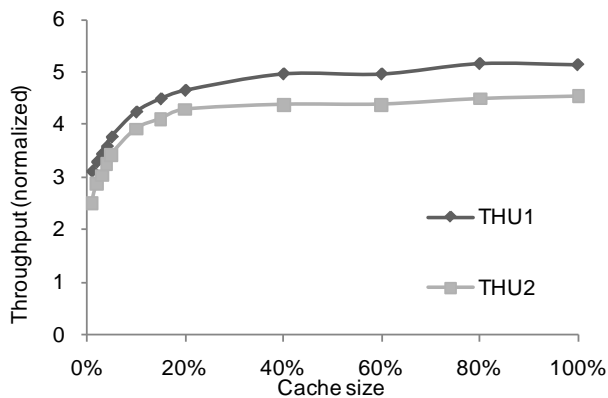

Figure 8. Throughput of proposed traffic classifier comparing to L7-filter.

When using our replacement algorithm, the cache hit ratio reaches almost its maximum when the cache size is 40% of total number of aggregate-flows. For most traces, the performance degradation is only around 10% if the cache size is reduced to 15% of total number of entries. Therefore, the performance of the traffic classifier can be guaranteed when the memory-usage of aggregate-flow cache is limited.

In Figure 8, the throughput of the proposed traffic classifier, normalized to that of L7-filter on the same traces, is illustrated. Only results on trace THU1 and THU2 are tested since these two traces are with packet payload, and the other traces are packet-header-only. The throughput of the traffic classifier increases up to 5.1 times over original L7-filter on trace THU1, up to 4.5 times on trace THU2. The increase is 4.5 times and 4.1 times respectively when the cache size is limited to 15% of total number of aggregate-flows.

## 6. CONCLUSIONS

In this paper, we study network application identification problem from a different perspective. We propose an application identification framework that classifies traffic at aggregate-flow level. Traffic classifier designer can leverage aggregate-flow heuristics provided by the aggregate-flow cache to achieve different design goal. A detailed traffic classifier designed based on the framework is demonstrated to improve the throughput of payload-based identification methods. Experiments on real-world traces show that our traffic classifier with can reduce up to 95% workload of backend identification engine. In addition, we optimize the classifier by proposing an effective design of aggregate-flow cache. We first characterize the temporal locality of aggregate-flow cache and propose a frequency based, recency-aware cache replacement algorithm based on the locality analysis. The proposed cache replacement algorithm outperforms well-known replacement algorithms, and achieves 90% of the optimal performance using only 15% of memory. Tests on an implementation of our classifier shows that the throughput of L7-filter is increased by up to 5.1 times by using our traffic classifier design.